\documentclass[prl,aps,twocolumn,showpacs]{revtex4}
\usepackage{epsfig}
\begin{document}
\title{Onset of thermal convection in a horizontal layer of granular gas}
\author{Evgeniy Khain and Baruch Meerson}
\affiliation{Racah Institute of Physics, Hebrew University of
Jerusalem, Jerusalem 91904, Israel}

\begin{abstract}
The Navier-Stokes granular hydrodynamics is employed for
determining the threshold of thermal convection in an infinite
horizontal layer of granular gas. The dependence of the convection
threshold, in terms of the inelasticity of particle collisions, on
the Froude and Knudsen numbers is found. A simple necessary
condition for convection is formulated in terms of the
Schwarzschild's criterion, well-known in thermal convection of
(compressible) classical fluids. The morphology of convection
cells at the onset is determined. At large Froude numbers, the
Froude number drops out of the problem. As the Froude number goes
to zero, the convection instability turns into a recently
discovered phase separation instability.
\end{abstract}

 \maketitle

\section{Introduction}

Fluidized granular media exhibit a plethora of fascinating
pattern-formation phenomena that have been subjects of much recent
interest \cite{patterngran}. In this work we address thermal
(buoyancy-driven) granular convection
\cite{Ramirez,Wildman,Sunthar,Meerson,Talbot}. Being unrelated to
the shear or time-dependence introduced by the system boundaries,
it resembles the Rayleigh-B\`{e}nard convection in classical fluid
\cite{Chandrasekhar} and its compressible modifications
\cite{Landau,Spiegel,GS,Gitterman,CU}. In classical fluid
convection requires an externally imposed negative temperature
gradient, that is a temperature gradient in the direction opposite
to gravity. In a vibrofluidized granular medium a negative
temperature gradient sets in spontaneously because of the energy
loss by inelastic collisions. Convection develops when the
absolute value of the temperature gradient is large enough. In the
simplest model of inelastic hard spheres that we will use it
happens when the inelasticity coefficient $q=(1-r)/2$ exceeds a
critical value depending on the rest of the parameters of the
system. Here $r$ is the coefficient of normal restitution of
particle collisions.

Thermal granular convection was first observed in molecular
dynamics (MD) simulations of a system of inelastically colliding
disks in a two-dimensional (2D) square box \cite{Ramirez}. The
boundaries of the box  \cite{Ramirez} did not introduce any shear
or time-dependence: the system was driven by a stress-free
thermalizing base. The top wall was perfectly elastic, while the
lateral boundaries were either elastic or periodic. Experiment
with a highly fluidized three-dimensional granular flow
\cite{Wildman} gives strong evidence for thermal convection,
though energy loss at the side walls introduces complications
\cite{Wildman,Talbot}. A clear identification of thermal
convection in experiment requires a large aspect ratio in the
horizontal direction, so that \textit{multiple} convection cells
can be observed. MD simulations in 2D of a vibrofluidized granular
system with a large aspect ratio indeed show multiple convection
cells \cite{Sunthar}.

This work deals with a theory of thermal granular convection in a
system with a large aspect ratio. Recently, a continuum model of
thermal granular convection has been formulated \cite{Meerson} in
the framework of the Navier-Stokes granular hydrodynamics. In the
dilute limit the Navier-Stokes hydrodynamics (or, more precisely,
gasdynamics) is systematically derivable from more fundamental
kinetic equations \cite{Jenkins,Brey1}. Like any other
hydrodynamic approach, the Navier-Stokes hydrodynamics demands
small Knudsen numbers for its validity. In addition, it has been
shown that at moderate inelasticities $q$ non-hydrodynamic effects
(such as the lack of scale separation, the normal stress
difference and non-Gaussianity in the particle velocity
distribution) may become important \cite{Goldhirsch}. Therefore,
the Navier-Stokes granular hydrodynamics is expected to be
accurate quantitatively only for nearly elastic collisions, $q \ll
1$. Though restrictive, the nearly elastic limit is conceptually
important. Also, one can expect some of the results, obtained in
this limit, to be still qualitatively valid for larger
inelasticities, such as those encountered in experiment.

In Ref. \cite{Meerson} the full set of nonlinear hydrodynamic
equations for thermal granular convection was solved numerically
in a 2D box with aspect ratio 1. It was observed, in qualitative
agreement with MD simulations \cite{Ramirez}, that the static
state of the system (a steady state with a zero mean flow) gives
way to convection via a supercritical bifurcation, the
inelasticity $q$ being the control parameter. The present work
employs the same hydrodynamic formulation \cite{Meerson} to
perform a systematic \textit{linear stability} analysis of the
static state. We determine the convection threshold as a function
of the scaled parameters of the problem and of the horizontal wave
number of small perturbations. This analysis makes it possible to
predict the convection threshold and determine the morphology of
the convection cells in a system with an arbitrary aspect ratio,
including an infinite horizontal layer, a standard setting for
convection in classical fluids \cite{Chandrasekhar,Cross}.  We
also formulate a simple \textit{necessary} (but not sufficient)
criterion for thermal granular convection in terms of the
Schwarzschild's criterion, well-known in thermal convection of
(compressible) classical fluids \cite{Landau}. Finally, we take
the limit of a zero gravity and establish the connection between
thermal convection and a recently discovered phase-separation
instability \cite{Livne,Brey,Khain,Livne2,Argentina,MPSS}.

\section{Model and static state}

Let a big number of identical smooth hard disks with diameter $d$
and mass $m$ move and inelastically collide inside an infinite
two-dimensional horizontal layer with height $H$. The gravity
acceleration $g$ is in the negative $y$ direction. The system is
driven by a rapidly vibrating base. We shall model it in a
simplified way by prescribing a constant granular temperature
$T_0$ at $y=0$. The top wall is assumed elastic. Hydrodynamics
deals with coarse-grained fields: the number density of grains
$n(\mathbf{r},t)$, granular temperature $T(\mathbf{r},t)$ and mean
flow velocity $\mathbf{v}(\mathbf{r},t)$. In the dilute limit, the
scaled governing equations are \cite{Meerson}:
\begin{equation}
d n / d t + n\, {\mathbf \nabla} \cdot {\mathbf v} = 0\,,
\label{cont}
\end{equation}
\begin{equation}
n \, d{\mathbf v}/dt = {\mathbf \nabla} \cdot {\mathbf P} - F \,
n\, {\mathbf e_y}\,, \label{momentum}
\end{equation}
\begin{equation}
n\, dT/dt + nT\, {\mathbf \nabla} \cdot {\mathbf v} = K\, {\mathbf
\nabla} \cdot (T^{1/2} {\mathbf \nabla} T) -K R\, n^{2}\,
   T^{3/2}.
\label{heat}
\end{equation}
Here  $d/dt = \partial/\partial t + {\mathbf v} \cdot {\mathbf
\nabla}$ is the total derivative, ${\mathbf P} = - nT \, {\mathbf
I} + (1/2)\, K\, T^{1/2}\, \hat{{\mathbf D}} \,$ is the stress
tensor, ${\mathbf D} = (1/2)\left({\mathbf \nabla v} + ({\mathbf
\nabla v})^{T}\right)$ is the rate of deformation tensor,
$\hat{{\mathbf D}}={\mathbf D}-(1/2)\, {\rm tr}\, ({\mathbf
D}\,)\, {\mathbf I}$ is the deviatoric part of ${\mathbf D}$, and
${\mathbf I}$ is the identity tensor. In the dilute limit, the
bulk viscosity of the gas is negligible compared to the shear
viscosity \cite{Jenkins}, so only the shear viscosity is taken
into account. In addition, we have neglected the small viscous
heating term in the heat balance equation (\ref{heat}). The
inelastic contribution to the heat flux \cite{Brey1} is
proportional to $q$ and can be safely neglected at small $q$. In
the 2D geometry, the three scaled parameters entering Eqs.
(\ref{momentum}) and (\ref{heat}) are the Froude number $F=m g
H/T_0$, the Knudsen number $K=2\pi^{-1/2}(d H \langle
n\rangle)^{-1} $ and the relative heat loss parameter
$R=8qK^{-2}$. Furthermore, $\langle n \rangle$ is the total number
of particles per unit length in the horizontal direction, divided
by the layer height $H$. It will be convenient to use the relative
heat loss number $R$ instead of $q$. In Eqs.
(\ref{cont})-(\ref{heat}), the distance is measured in the units
of $H$, the time in units of $H/T_0^{1/2}$, the density in units
of $\langle n \rangle$, the temperature in units of $T_0$, and the
velocity in units of $T_0^{1/2}$. The Navier-Stokes hydrodynamic
model is expected to be valid when the mean free path of the
particles is much smaller than any length scale (and the mean
collision time is much smaller than any time scale) described
hydrodynamically. This implies, in particular, that the Knudsen
number should be small: $K \ll 1$.

The boundary conditions for the temperature are $T(x, y=0, t)=1$
at the base and a zero normal component of the heat flux at the
upper wall: $\partial T/\partial y(x, y=1, t)=0$. For velocities
we demand zero normal components and slip (no stress) conditions
at the boundaries. The total number of particles in the system is
conserved:
\begin{equation}
\lim_{L\rightarrow \infty}\frac{1}{2L}\int_{-L}^{L} dx\int_0^1
dy\, n(x,y,t) = 1\,, \label{conservation}
\end{equation}
where we have introduced the horizontal dimension $2L$. The
hydrodynamic problem is characterized by the three scaled numbers
$F$, $K$ and $R$.

\begin{figure}[ht]
\vspace{-0.3 cm} \center{\epsfxsize=6.0 cm 
\epsffile{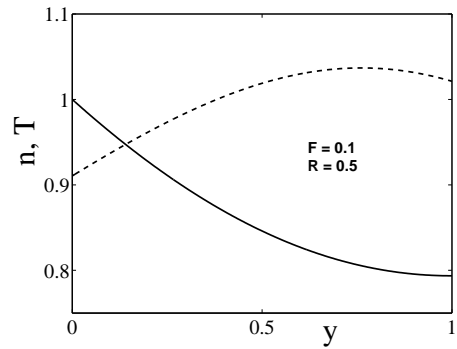}} \caption{Static temperature (solid line) and
density (dashed line) profiles for $F=0.1$ and $R=0.5$.}
\label{steady1}
\end{figure}

The simplest steady state of the system is the static state: no
mean flow. At a nonzero gravity the density $n_s$ and temperature
$T_s$ of the static state depend only on $y$, and are described by
the equations
\begin{equation}
(n_s\,T_s)^{\prime} + F \,n_s = 0 \,\,\,\mbox{and}\,\,\,
(T_s^{1/2} T_s^{\prime})^{\prime} - R\, n_s^{2}\, T_s^{3/2} = 0\,,
\label{steady}
\end{equation}
where the primes denote the $y$-derivatives. The boundary
conditions are $T_s(0)=1$ and $T_s^{\prime}\, (1) = 0$, the
normalization condition is $ \int_0^1 dy\, n_s (y)=1\,. $ The
static state (see Fig. \ref{steady1}) is characterized by {\it
two} scaled numbers: $F$ and $R$, and can be found analytically by
transforming from the Eulerian coordinate $y$ to the Lagrangian
mass coordinate $\mu(y)=\int_0^y n_{s}(y^{\prime})dy^{\prime}$
\cite{Meerson}. At large enough $R$, a density inversion develops:
a denser (heavier) gas is located on the top of an under-dense
(lighter) gas. This is clearly a destabilizing effect that drives
thermal convection. However, this effect is neither sufficient,
nor necessary for convection, see below. The actual (necessary and
sufficient) criterion should take into account the heat conduction
and viscosity that scale like $K$ in the governing equations and
have a stabilizing role.  A small sinusoidal perturbation in the
horizontal direction is unstable with respect to convection if the
relative heat loss parameter $R$ exceeds a critical value that
depends on $F$, $K$ and the horizontal wave number.

\section{The linear stability analysis}

The linear stability analysis involves linearization of Eqs.
(\ref{cont})-(\ref{heat}) around the static solutions $n_s(y)$ and
$T_s(y)$.  The linearized equations are
\begin{eqnarray}
\frac{\partial \tilde{n}}{\partial t} + n_s \frac{\partial
v_x}{\partial x} + \frac{\partial}{\partial y} (n_s v_y) = 0\,,
 \label{cont1}
\end{eqnarray}
\begin{eqnarray}
n_s \frac{\partial {\mathbf v}}{\partial t} = -{\mathbf \nabla}
(n_s \tilde{T}+T_s \tilde{n}) + \frac{1}{2} K {\mathbf \nabla}
\cdot (T_s^{1/2}\hat{{\mathbf D}}) \nonumber \\ - F \tilde{n}\,
{\mathbf e_y}, \label{momentum1}
\end{eqnarray}
\begin{eqnarray}
n_s \left(\frac{\partial \tilde{T}}{\partial t} + T_s^{\prime} v_y
\right) +n_s T_s {\mathbf \nabla} \cdot {\mathbf v} = \nonumber
\\ K \nabla^2 (T_s^{1/2} \tilde{T})  - K R\, n_s^2 T_s^{3/2}
\left(\frac{2\tilde{n}}{n_s} + \frac{3\tilde{T}}{2T_s} \right)\,,
\label{heat1}
\end{eqnarray}
where $\tilde{n}$, $\tilde{T}$ and ${\mathbf v}$ denote small
perturbations. Exploiting the translational symmetry  of the
static state in the horizontal direction, one can consider a
single Fourier mode in $x$:
\begin{eqnarray}
\tilde{n}(x,y,t)= {\rm e}^{-\gamma t}\, N(y) \cos k_{x}x\,,
\nonumber
\\ \tilde{T}(x,y,t)= {\rm e}^{-\gamma t}\, \Theta (y) \cos k_{x}x\,, \nonumber
\\ v_x(x,y,t)= {\rm e}^{-\gamma t}\, u(y) \sin k_{x}x\,, \nonumber \\
v_y(x,y,t)= {\rm e}^{-\gamma t}\, v(y) \cos k_{x}x\,,
\label{modes}
\end{eqnarray}
where $\gamma$ is the scaled growth/decay rate, and $k_x$ is the
scaled horizontal wave number. Substituting Eq. (\ref{modes}) into
Eqs. (\ref{cont1})-(\ref{heat1}) and eliminating $N$, we obtain
three homogeneous ordinary differential equations that can be
written as a single equation for the eigenvector ${\mathbf U} (y)
= \left[ \Theta(y),\, u(y),\,v(y) \right]$, corresponding to the
eigenvalue $\gamma$:
\begin{equation}
{\mathbf A}\, {\mathbf U}^{\prime \prime} + {\mathbf B}\, {\mathbf
U}^{\prime} + {\mathbf C}\, {\mathbf U} = 0\,,
\label{vectorsystem}
\end{equation}
where
\begin{equation}
{\mathbf A} = \left(
\begin{array}{ccc}
K\,a_0 & 0                & 0        \\ 0    &  K\,a_0 / 4 & 0
\\ 0    & 0                & K\,a_0 /4 - n_s
T_s / \gamma
\end{array} \right)\,,
\label{A}
\end{equation}
\begin{equation}
{\mathbf B} = \left(
\begin{array}{ccc}
2K a_1 & 0 & -2KR n_s^2 T_s^{3/2} / \gamma - n_s T_s\ \\

0 & K\, a_1 / 4 & k_{x} n_s T_s / \gamma      \\

 -n_s &  - k_{x} n_s T_s / \gamma & K\,a_1 /4 -T_s n_s^{\prime} / \gamma
\end{array} \right)\,,
\label{B}
\end{equation}
while the elements of matrix ${\mathbf C}$ are
\begin{eqnarray}
C_{11} &=& K\,a_2 - 3 K R n_s^2 a_0 / 2 - K k_{x}^2 a_0 +\gamma
n_s\,,\nonumber
\\
C_{12} &=& - 2 K R n_s^2 T_s^{3/2} k_{x} / \gamma - k_{x} n_s
T_s\,,\nonumber
\\
C_{13} &=& - 2 K R n_s T_s^{3/2} n_s^{\prime} / \gamma - n_s
 T_s^{\prime}\,,\nonumber
\\
C_{21} &=& k_{x} n_s\,,\nonumber
\\
C_{22} &=& k_{x}^2 n_s T_s / \gamma + \gamma n_s - K k_{x}^2 a_0 /
4 \,,\nonumber
\\
C_{23} &=& k_{x} T_s n_s^{\prime} / \gamma - K k_{x} a_1 / 4
\,,\nonumber
\\
C_{31} &=& - n_s^{\prime}\,,\nonumber
\\
C_{32} &=& - K k_{x} a_1 / 4\,,\nonumber
\\
C_{33}&=&-(T_s^{\prime\prime}n_s+T_s^{\prime}n_s^{\prime}) /
\gamma + \gamma n_s - K k_{x}^2 a_0 / 4\,. \label{C}
\end{eqnarray}
We have denoted for brevity $a_0 = T_s^{1/2}(y)\,,\,  a_1 =
a_0^{\prime}$ and $a_2 = a_0^{\prime\prime}$. The boundary
conditions for the functions $\Theta$, $u$ and $v$ are the
following:
\begin{equation}
\Theta (0)=\Theta^{\prime} (1) = u^{\prime} (0) = u^{\prime} (1) = v (0) = v (1) = 0\,.
\label{boundcond}
\end{equation}
Equation (\ref{vectorsystem}) with the boundary conditions
(\ref{boundcond}) define a linear \textit{boundary-value} problem:
there are three boundary conditions at the base and three at the
top.  A simple numerical procedure (realized in MATLAB) enabled us
to avoid the unpleasant shooting in three parameters. The
procedure employs the linearity of the problem. We first
complement the three boundary conditions at the base by three
\textit{arbitrary} boundary conditions \textit{at the base}, and
compute numerically three independent solutions of Eqs.
(\ref{vectorsystem}). The \textit{general} solution can be
represented as a linear combination of these three independent
solutions that includes three arbitrary coefficients. Demanding
that the three remaining boundary conditions \textit{at the top}
be satisfied, we obtain three homogeneous linear algebraic
equations for the coefficients. A nontrivial solution requires
that the determinant vanish, which yields the eigenvalue $\gamma$.
Varying $R$ at fixed $F$, $K$ and $k_x$, we determine the critical
value $R=R_c$ for instability from the condition ${\rm Re}
\,\gamma = 0$. We found this algorithm to be accurate and
efficient.

\begin{figure}[ht]
\begin{tabular}{cc}
\vspace{-0.3 cm}
\epsfysize=5 cm 
\epsffile{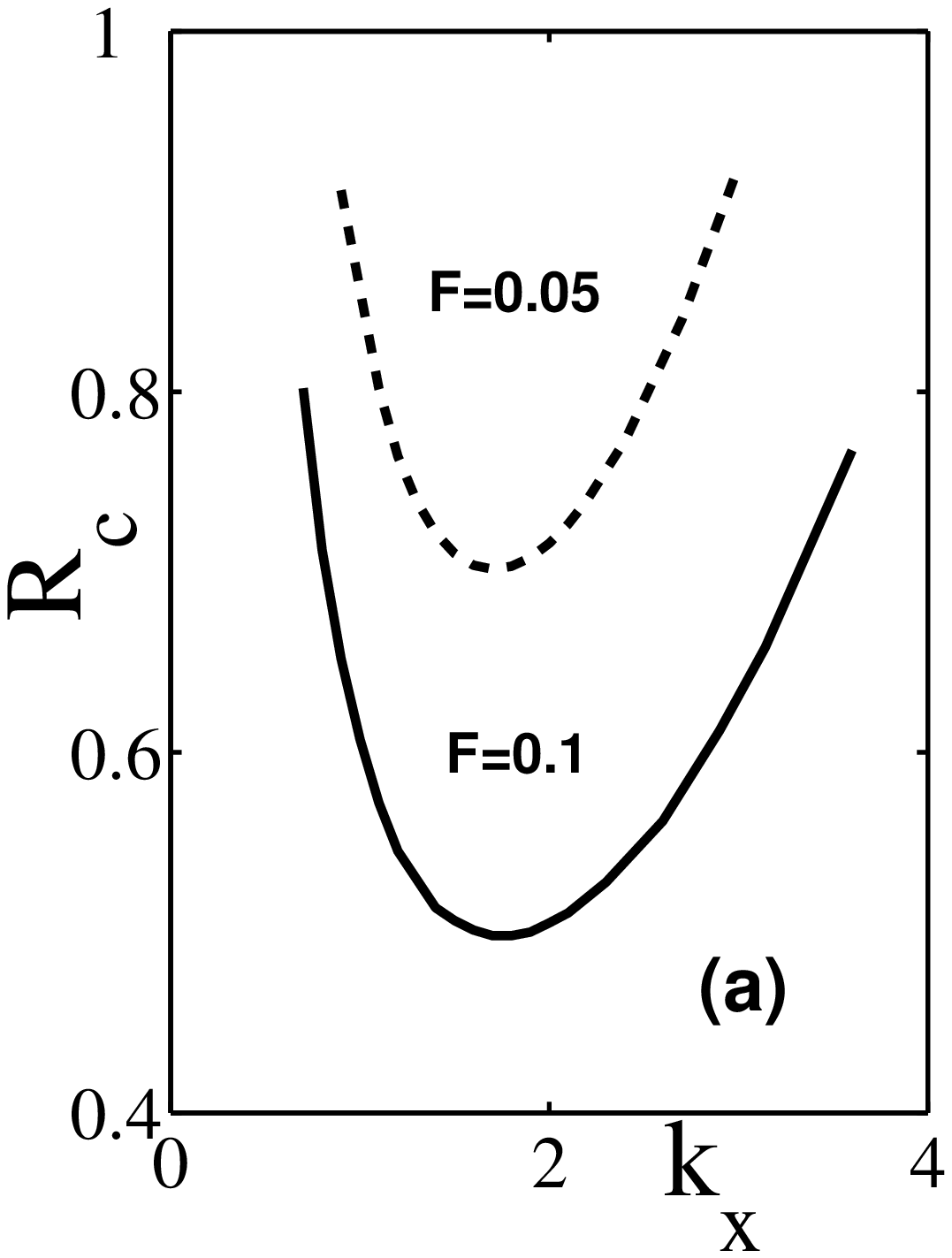} &\hspace{0.2cm}
\epsfysize=5 cm 
\epsffile{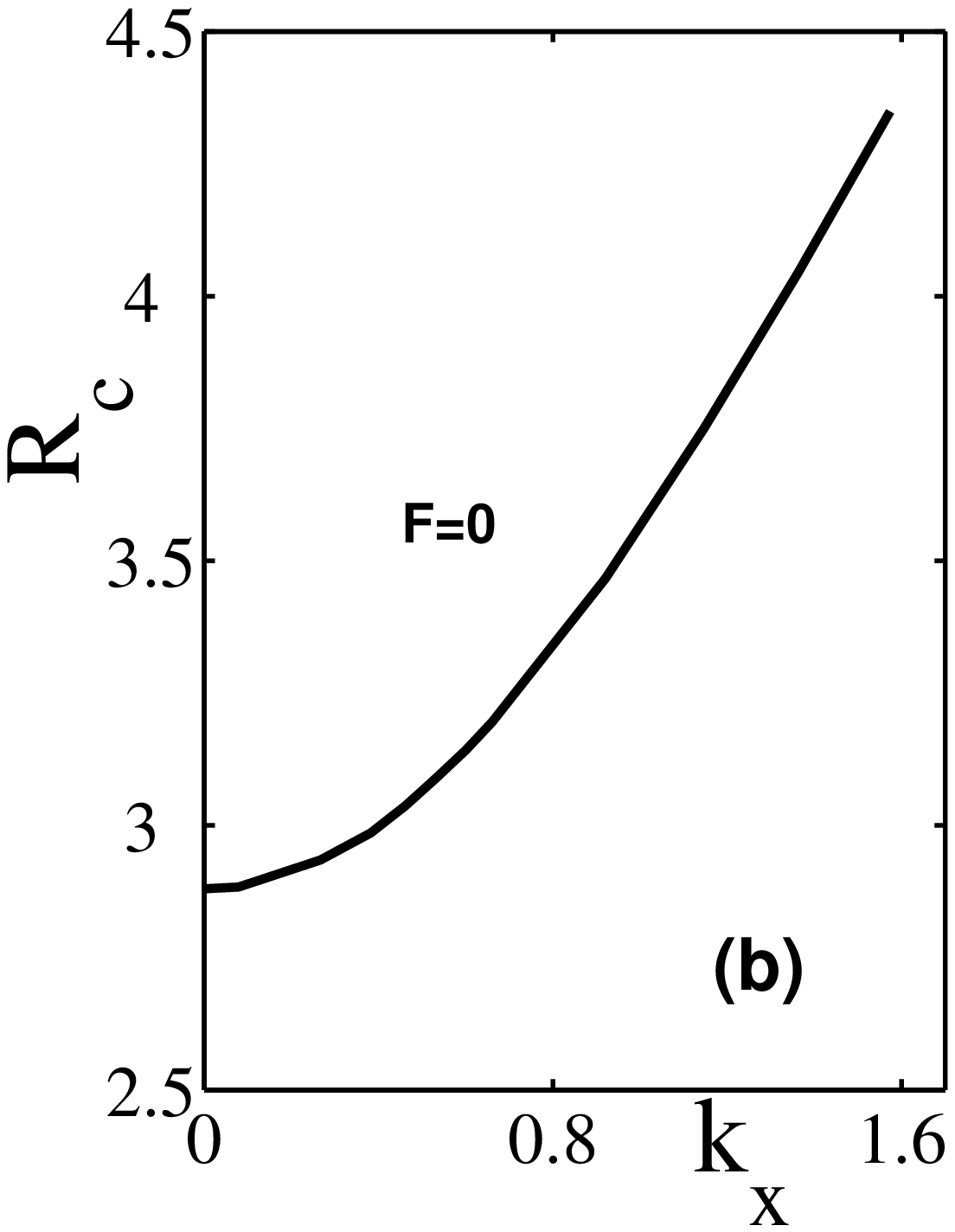}
\end{tabular}
\caption{The critical values of the relative heat loss parameter
$R$ for the convection instability (non-zero gravity, a) and
phase-separation instability (zero gravity, b) versus the
horizontal wave number $k_{x}$. The Knudsen number $K=0.02$.}
\label{Rkx}
\end{figure}

\begin{figure}[ht]
\begin{tabular}{cc}
\epsfysize=2.2 cm 
\epsffile{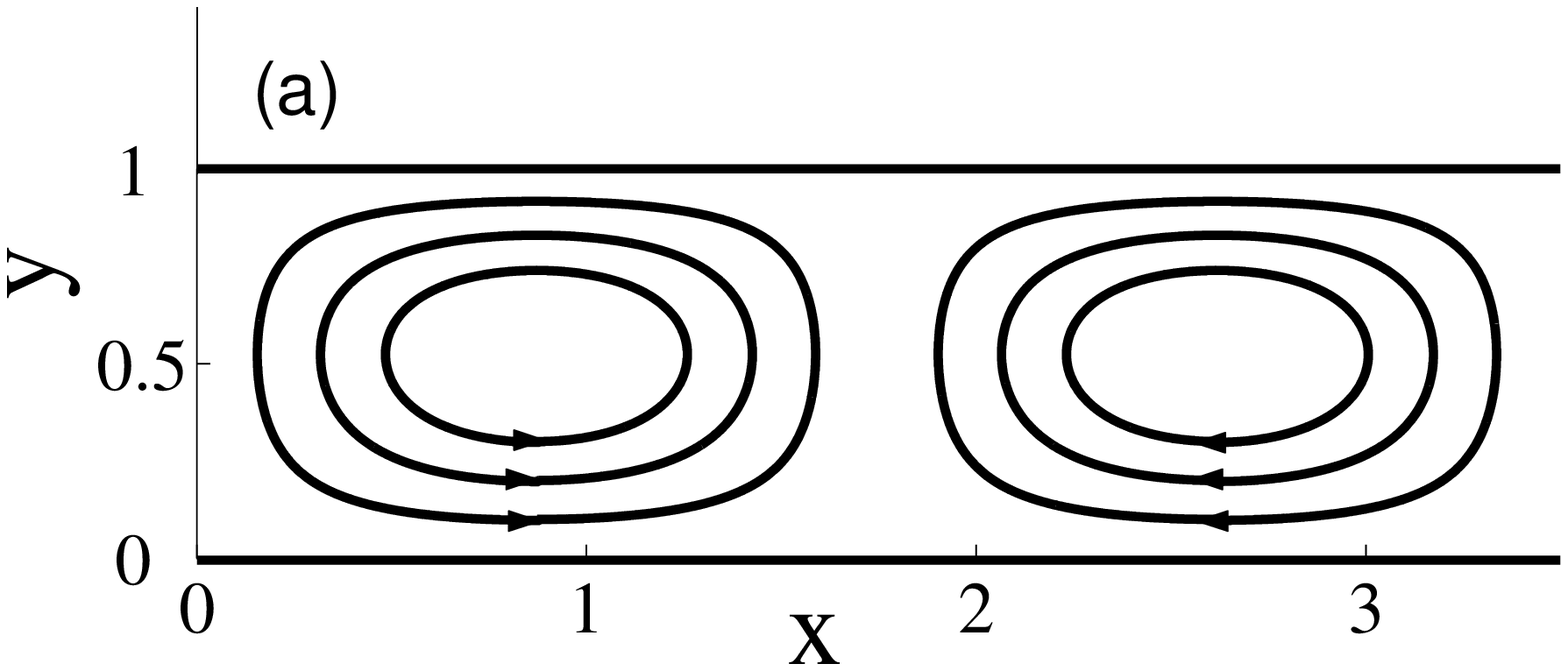} &\hspace{-0.2cm} \vspace{-0.3 cm}
\epsfysize=2.1 cm 
\epsffile{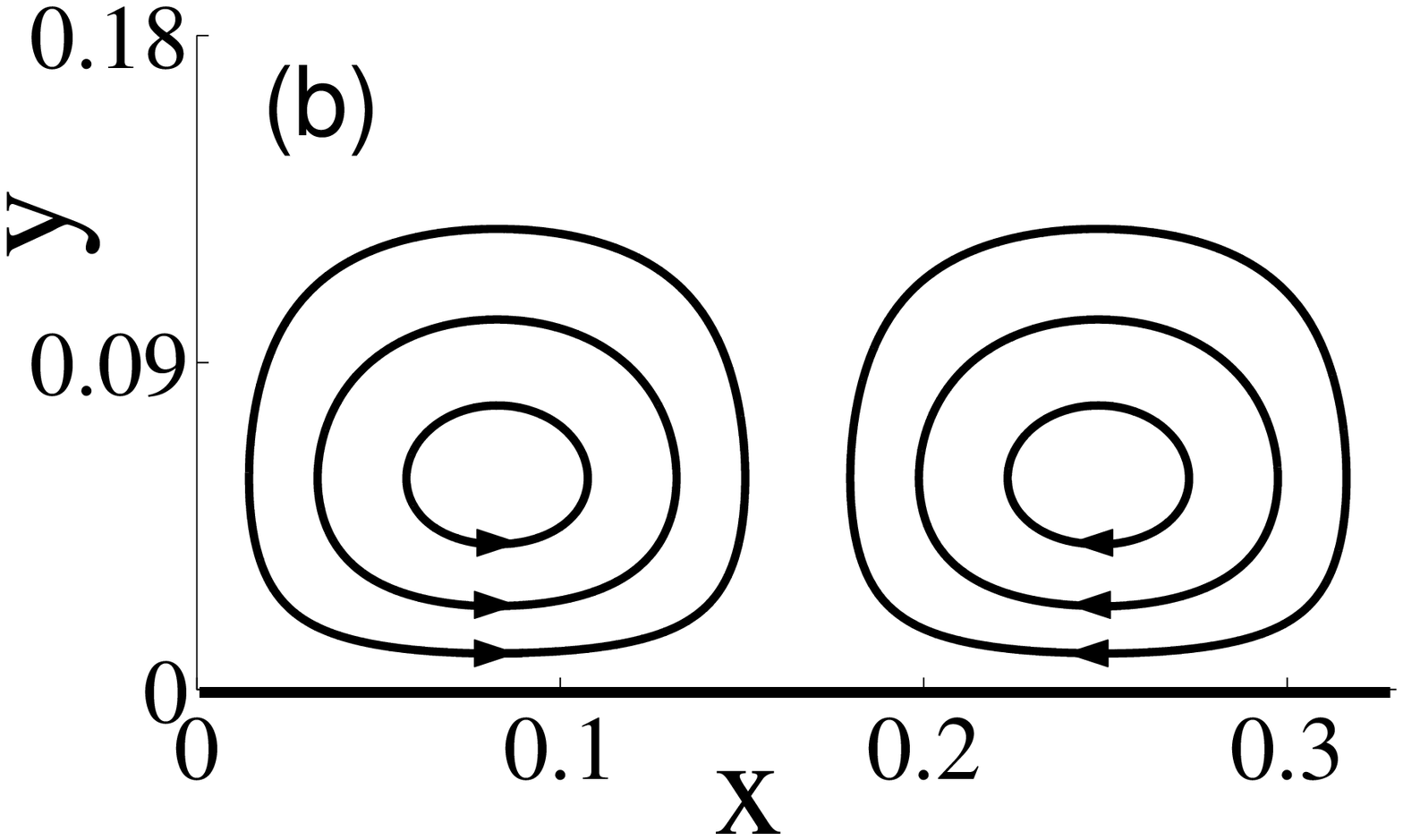}
\end{tabular}
\caption{The convection cells at the instability onset
$k=k_{x}^{*}$ and $R=R_{c}^{*}$.  The Knudsen number is $K=0.02$.
Figure a corresponds to the solid curve of Fig. \ref{Rkx}a. Here
$F=0.1$, $k_{x}^{*}=1.8$ and $R_{c}^{*}=0.49$. Figure b
corresponds to the large-$F$ limit, when the  granulate is
localized at the base. Here $F=5$, $k_{x}^{*}=19$, and
$R_{c}^{*}=2.86$.} \label{cell}
\end{figure}

In the whole region of the parameter space that we explored we
found that ${\rm Im} \, \gamma =0$ at the instability onset.
Therefore, thermal granular convection does not exhibit
overstability and can be analyzed in terms of marginal stability.
Figure \ref{Rkx}a shows the marginal stability curves versus the
horizontal wave number $k_x$ at a fixed $K$ and two different
values of $F$ \cite{comparison}. The curves exhibit minima
$(k_{x}^{*}, R_{c}^{*})$, similarly to the convection in classical
fluids \cite{Chandrasekhar}. Therefore, the convection threshold
in the horizontally infinite layer is $R=R_{c}^{*}$. Close to the
onset, the expected horizontal size of the convection cell is $
2\pi/k_{x}^{*}$. Figure \ref{cell} depicts these convection cells
obtained by plotting the field lines of the respective velocity
field $\left(u(y) \sin k_{x}x\,, \, v(y) \cos k_{x}x\right),$
found numerically. One can see that at small $F$ the cells occupy
the whole layer of granular gas and are elongated in the
horizontal direction. At large $F$ the cells are effectively
located near the base, and their aspect ratio is close to unity.

Figure 2b corresponds to a zero gravity: $F=0$. Here a different
symmetry-breaking instability occurs: the one that leads to
\textit{phase separation}
\cite{Livne,Brey,Khain,Livne2,Argentina,MPSS}. When $R$ exceeds
the marginal stability threshold $R_{c}^{*}(F=0)$, the laterally
symmetric stripe of enhanced particle density at the top wall
becomes unstable and gives way to a 2D steady state. In contrast
to the convection, the new steady state with a broken
translational symmetry is \textit{static}: no mean flow. The
quantity $R_{c}^{*}(F=0)$ can be calculated analytically
\cite{Khain}. In our present notation, it is determined from the
algebraic equation $\coth \mu = \mu$, where $\mu=
(R_{c}^{*}/2)^{1/2}$. This yields $R_{c}^{*}(F=0)=2.8785\dots$.
The minimum of the marginal stability curve occurs here at
$k_{x}^{*}=0$, that is for an infinitely long wavelength.

\begin{figure}[ht]
\vspace{-0.3 cm} \center{\epsfxsize=6.0 cm 
\epsffile{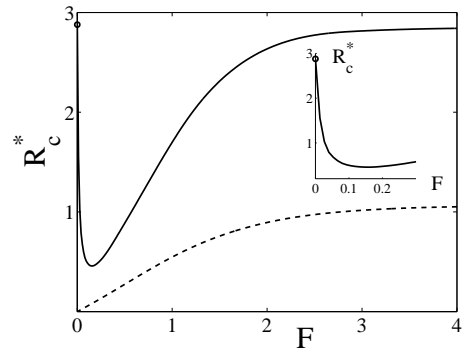}} \caption{The convection threshold versus the
Froude number $F$. The solid line shows the $R_{c}(F)$ curve at
$K=0.02$. At large $F$ the threshold $R_{c}^{*}$ approaches
$2.85...$. The dashed line is the Schwarzschild's curve $R_{S}(F)$
that gives a \textit{necessary} condition for convection. The
small-$F$ asymptotics of the Schwarzschild's curve is $R_{S}\simeq
F/2$. In the limit of large $F$, $R_{S}(F)$ approaches
$1.06514\dots$. The inset shows that, as $F\to 0$, $R_{c}^{*}$
approaches $2.8785\dots$, the threshold of the phase-separation
instability.} \label{RcF}
\end{figure}

We found that the crossover between the two instabilities is
continuous.  The dependences of $R_{c}^{*}$ and $k_{x}^{*}$ on the
Froude number $F$ are shown in Fig. \ref{RcF}. One can see that
the $R_{c}^{*} (F)$ dependence is non-monotonic. A stronger
gravity is favorable for convection at very small $F$ (as Fig.
\ref{Rkx}a also shows). However, this tendency is reversed at $F
\simeq 0.16$, and $R_c^{*}$ starts to grow with $F$ until it
saturates at large $F$. In its turn, $k_{x}^{*}$ goes down
monotonically with $F$ and vanishes at $F=0$. The decrease is
quite slow at intermediate $F$, but becomes very rapid at very
small $F$. As the phase separation instability does not exist in
classical fluid, this low-$F$ behavior is unique for granular
fluid.

\begin{figure}[ht]
\vspace{-0.3 cm} \center{\epsfxsize=6.0 cm 
\epsffile{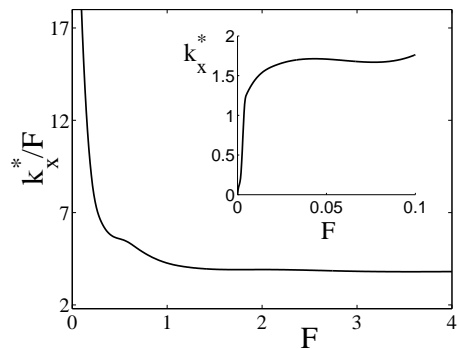}} \caption{The convection threshold versus the
Froude number $F$. Shown is the $k_{x}^{*}(F)$  dependence at
$K=0.02$. The inset shows that $k_{x}^{*}$ goes to zero as $F \to
0$. For large $F$, $k_{x}^{*}/F \to 3.80\dots$.} \label{kxF}
\end{figure}

The large-$F$ limit deserves a special attention. Here the
granulate is localized at the base. This regime is convenient in
experiment, as particle collisions with the top wall (which are in
reality inelastic) are avoided. A natural unit of distance in this
regime is $\lambda=T_0/mg$, while the time should be scaled to
$\lambda/T_0^{1/2}$. Correspondingly, $\langle n \rangle$ is
defined now as the total number of particles per unit length in
the horizontal direction, divided by $\lambda$. After rescaling
the Froude number $F$ drops out of Eqs. (\ref{cont})-(\ref{heat})
and enters the problem only via the top wall position
$H/\lambda=F$. For $F \gg 1$ the top wall can be safely moved to
infinity, and $F$ drops out of the problem completely. Therefore,
at large $F$, the convection threshold $R_{c}^{*}$ should depend
only on $K$. Our numerical results fully support this prediction,
see Fig. \ref{RcF}. How should $k_{x}^{*}$ behave at large $F$?
Let us reintroduce for a moment the ``physical'' (dimensional)
horizontal wave number $k_{x\,ph}$. The scaled critical wave
number $k_{x}^{*}= k_{x\,ph}^{*} H= k_{x\,ph}^{*} \lambda \, F$.
As the product $k_{x\,ph}^{*} \lambda$ is the scaled wave number
in the newly rescaled variables, it should be independent of $F$
at large $F$. Therefore, $k_{x}^{*}$ should be proportional to
$F$. Figure \ref{kxF} shows that the quantity $k_{x}^{*}/F$ indeed
approaches a constant (that depends on $K$) at large $F$.

These results give, for fixed values of $F$ and $K$, the necessary
and sufficient criterion for convection. It is often useful to
also have a simpler and easier-to-interpret criterion, even if
approximate. A simplified criterion for convection can be obtained
by neglecting the viscosity and heat conduction terms in the
linearized equations (\ref{cont1})-(\ref{heat1}), that is by
taking the limit $K \to 0$. As the viscosity and heat conduction
act against convection, this procedure obviously yields a
\textit{necessary}, but not sufficient, criterion for convection.

Without the dissipative terms, Eqs. (\ref{cont1})-(\ref{heat1})
coincide with the linearized equations of \textit{ideal}
hydrodynamics of \textit{classical} fluid with specified static
profiles of temperature $T_s(y)$ and density $n_s(y)$. Even for
this idealized problem, the exact criterion for convection can be
obtained only numerically, and the result depends explicitly on
the specific profiles $T_s(y)$ and $n_s(y)$. There is, however, a
simple and general limit here in terms of the Schwarzschild's
criterion \cite{Landau} that yields a lower bound for the
convection threshold. The Schwarzschild's criterion guarantees
that there is \textit{no} convection if the entropy of the fluid
in the static state $S(n_s,T_s)$ grows with the height, that is
$S^{\prime}(n_s,T_s)>0$ for \textit{any} $y$. For the nearly
elastic hard sphere model in 2D the granular entropy in the dilute
limit is $S(n,T)=\ln (T/n)$. [Importantly, we are not making here
any additional assumption: this simple constitutive relation for
$S(n,T)$ was already used in Eqs. (\ref{cont})-(\ref{heat}).]
Therefore, the Schwarzschild's criterion can be rewritten in terms
of the static temperature and density profiles and their first
derivatives. For a given $F$, the static profiles are determined
solely by the relative heat loss parameter $R$. Therefore, the
Schwarzschild's criterion yields a critical value $R=R_S(F)$ so
that at $R<R_S(F)$ there is \textit{no} convection. The opposite
inequality $R>R_S(F)$ yields a \textit{necessary} (but of course
not sufficient) criterion for convection. How to find $R_S(F)$? At
small enough $R$ the spatial derivative of the entropy,
$S^{\prime} (y)$, is positive at any height $y\ge 0$. By
increasing $R$ we observe that,  at the critical value $R=R_S(F)$,
the entropy derivative $S^{\prime} (y)$ vanishes at some point
$y$. It is crucial that this point is always $y=0$. Increasing $R$
further, we would already have an \textit{interval} of heights
where $S^{\prime}(y)<0$. Therefore, \textit{the Schwarzschild's
curve} $R_{S}(F)$ can be obtained from the condition
$S^{\prime}(y=0)=0$. This curve, obtained numerically, is shown by
the dashed line in Fig. \ref{RcF}. As expected, the exact
(necessary and sufficient) convection threshold curve always lies
\textit{above} the Schwarzschild's curve.

The small-$F$  and large-$F$ asymptotics of the Schwarzschild's
curve can be obtained analytically. Let us first consider the case
of $F \ll 1$. As will be seen from the result, here $R \ll 1$ too,
and one can represent the steady-state solutions as
$T_s(y)=1+\delta T_s(y)$, and $n_s(y)=1+\delta n_s(y)$, where
$\delta T_s \ll 1$ and $\delta n_s \ll 1$. Substituting these
expressions into Eq. (\ref{steady}) and keeping only the first
order quantities, we obtain two very simple linear differential
equations. Solving them with the respective boundary and
normalization conditions, we obtain $T_s(y)\simeq 1-Ry+(R/2)y^2$
and $n_s(y) \simeq 1+F/2-R/3+(R-F)y-(R/2)y^2$. The condition
$S^{\prime}(y=0)=0$ then yields the desired small-$F$ asymptotics:
$R_{S} (F \ll 1) \simeq F/2$.

At $F \gg 1$, one can conveniently use the analytic solution for
the static profiles in the Lagrangian mass coordinate
\cite{Meerson}. In this limit
\begin{equation}
T_{s}(z)=\frac{I_{0}^{2}(z)}{I_{0}^{2}(\sqrt{R/2}\,)}
\,,\,\,\,\,\,n_{s}(z)=\frac{I_{0}^{2}(\sqrt{R/2}\,)\,z}{\sqrt{R/2}\,I_{0}^{2}(z)}\,,
\label{profiles} \end{equation} where $z=(R/2)^{1/2}(1-\mu)$, and
$I_{n}(\dots)$ is the modified Bessel function of the first kind.
(Recall that here we rescale the Eulerian coordinate $y$ by
$\lambda=T_0/mg$ and define $\langle n \rangle$ as the number of
particles per unit length in the horizontal direction divided by
$\lambda$.) Using Eqs. (\ref{profiles}), we compute the
$y$-derivative of the entropy:
\begin{equation}
S^{\prime} = \frac{I_0^2(\sqrt{R/2}\,)}{I_0^3(z)}\,\left[I_0(z)-4
z I_1 (z)\right]\,. \label{S'}
\end{equation}
This expression vanishes when $I_0(z)-4 z I_1 (z)=0$ which occurs
at $z=z_*=0.72977\dots$. As $y=0$ corresponds to $\mu=0$, we
immediately obtain $R_S = 2 z_*^2 = 1.06514\dots$. At $R<R_S$ we
have $S^{\prime}>0$ everywhere, so $R=R_S$ indeed corresponds to
the Schwarzschild's criterion. The value of $R_S (F \gg 1)
=1.06514\dots$ shows up as the large-$F$ plateau of the
Schwarzschild's curve in Fig. \ref{RcF}.

\begin{figure}[ht]
\epsfig{file=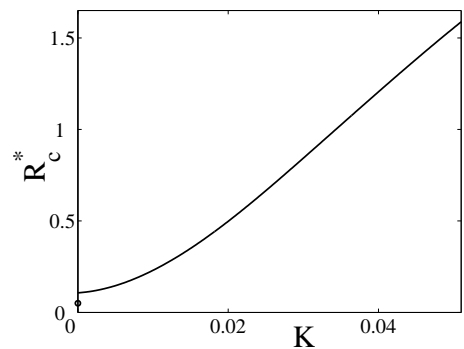, width=6.0cm, clip=} \caption{Convection
threshold $R_{c}^{*}$ versus the Knudsen number $K$ at $F=0.1$. As
$K \to 0$, the value of $R_{c}^{*}$ is greater than the
Schwarzschild's value $R_{S}(F=0.1) \simeq 0.05$, shown by the
empty circle.} \label{RcK}
\end{figure}

Now let us return to the exact (necessary and sufficient)
criterion for convection, found by solving the full linearized
problem numerically. The dependence of the convection threshold
$R_{c}^{*}$ on the Knudsen number $K$ (at a fixed $F$) is shown in
Fig. \ref{RcK}. $R_{c}^{*}$ grows with $K$, because the viscosity
and heat conduction (both of which scale like $K$) tend to
suppress convection \cite{Meerson}. As $K \to 0 $ the viscosity
and heat conduction become negligible. Still, one should expect
that the limiting value of the critical heat loss parameter $R_{K
\to 0}(F)=\lim_{K\rightarrow 0} R_{c}^{*}(F,K)$ is greater than
the Schwarzschild's lower bound $R_{S}(F)$. This is indeed what is
seen in Fig. \ref{RcK}, where the Schwarzschild's value
$R_{S}(F=0.1) \simeq 0.05$ is shown by the empty circle.

\section{Summary}

We performed a linear stability analysis of the static state in a
horizontal layer of granular gas driven from below. The
hydrodynamic theory \cite{Meerson} that we employed in our
analysis is expected to be valid when the mean free path of the
particles is much smaller than any length scale described
hydrodynamically. We have found the convection threshold, in terms
of the relative heat loss number $R$, versus the two other scaled
numbers of the problem: the Froude number $F$ and the Knudsen
number $K$. We have predicted the morphology of convection cells
at the onset of convection. As $F\to 0$, the convection
instability goes over continuously into the phase-separation
instability \cite{Livne,Brey,Khain,Livne2,Argentina,MPSS}. At
large $F$ the convection threshold depends only on $K$. We
established a simple connection between thermal granular
convection and classical thermal convection of ideal compressible
fluid. The connection is given in terms of the Schwarzschild's
criterion, a universal necessary (but not sufficient) condition
for thermal convection. A further development of the theory should
account for the excluded-volume effects \cite{Jenkins}.
Importantly, the simple Schwarzschild's criterion will be readily
available in the finite-density theory. Indeed, this criterion
requires only the knowledge of the static profiles of the granular
temperature and density and the constitutive relation for the
granular entropy.

\section{Acknowledgments}

We gratefully acknowledge useful discussions with Igor S. Aranson,
John M. Finn, Xiaoyi He, Pavel V. Sasorov and Victor Steinberg.
The work was supported by the Israel Science Foundation
administered by the Israel Academy of Sciences and Humanities.

\end{document}